\documentclass[preprint2]{aastex6}

\def\COi{$^{12}$CO}
\def\COii{$^{13}$CO}
\def\COiii{C$^{18}$O}
\def\H2{H$_{2}$}
\def\HI{H{\scriptsize~I}}
\def\deg{$^{\circ}$}
\def\Msun{M$_{\odot}$}
\def\kms{km s$^{-1}$}
\def\Tex{$T_{\mathrm{ex}}$}
\def\sq{$^{2}$}
\def\Vlsr{$V_{\mathrm{LSR}}$}
\def\NH2{$N_{\mathrm{H_{2}}}$}

\usepackage{multirow}
\usepackage{color}
\usepackage{graphicx}

\usepackage{array}
\newcommand{\PreserveBackslash}[1]{\let\temp=\\#1\let\\=\temp}
\newcolumntype{C}[1]{>{\PreserveBackslash\centering}p{#1}}
\newcolumntype{R}[1]{>{\PreserveBackslash\raggedleft}p{#1}}
\newcolumntype{L}[1]{>{\PreserveBackslash\raggedright}p{#1}}

\begin{document}

\title{The Molecular Structures of the Local Arm and Perseus Arm \\
in the Galactic Region of
l = [139\deg.75, 149\deg.75] , b = [$-$5\deg.25, 5\deg.25] }

\author{Xinyu Du\altaffilmark{1,2,3},Ye Xu\altaffilmark{1,3}, Ji Yang\altaffilmark{1,3}, Yan Sun\altaffilmark{1,3}}
\affil{xydu@pmo.ac.cn; xuye@pmo.ac.cn}
\altaffiltext{1}{Purple Mountain Observatory, Chinese Academy of Science, Nanjing 210008, China}
\altaffiltext{2}{Graduate University of the Chinese Academy of Sciences, 19A Yuquan Road, Shijingshan District,
Beijing 100049, China}
\altaffiltext{3}{Key Laboratory of Radio Astronomy, Chinese Academy of Science, Nanjing 210008, China}

\begin{abstract}
Using the Purple Mountain Observatory Delingha (PMODLH) 13.7 m telescope,
we report a 96-square-degree \COi/\COii/\COiii\ mapping observation
toward the Galactic region of $l=[139.75,149.75]^{\circ}$, $b=[-5.25,5.25]^{\circ}$.
The molecular structures of Local Arm and Perseus Arm are presented.
Combining \HI\ data and part of the Outer Arm results,
we obtain that the warp structure of both atom and molecular gas is obvious,
while the flare structure only exists in atomic gas in this observing region.
In addition, five filamentary giant molecular clouds on the Perseus Arm are identified.
Among them, four are newly identified.
Their relations with Milky Way large scale structure are discussed.

\end{abstract}

\section{Introduction} \label{sec:intro}
The Milky Way (MW) galaxy is our home galaxy.
Since we are located at the Galactic plane, which is full of gas and dust,
the difficulty in studying its structure is greater than that for some external galaxies.
Up to now, although many previous works have already revealed its global appearance
(e.g., \citealt{Bok1959Obs}; \citealt{Oort1958MNRAS}; \citealt{GG1976A&A}),
its detailed structure is still under debate
(e.g., \citealt{Russeil2007A&A}; \citealt{Vallee2008AJ}; \citealt{Reid2014ApJ}).
For researching the Milk Way structure, the most basic requirement is the high-quality and large-scale survey data.
In the past several decades,
lots of molecular line surveys, especially the CO line surveys have been conducted
(cf. \citealt{HD2015ARA&A}).
Among them, two most remarkable CO surveys conducted by \citet{Heyer1998ApJS} and \citet{Dame2001ApJ}
have provided a lot of precious data.
Currently, many MW models are largely based on those two surveys
(e.g., \citealt{HH2014A&A}).
However, such data with relatively low resolution or low sensitivity gradually become insufficient
for more detailed research.
Fortunately, the on-going Milky Way Imaging Scroll Painting (MWISP) project
\footnote{http://english.dlh.pmo.cas.cn/ic/ \\ or http://www.radioast.nsdc.cn/mwisp.php;
One can submit an application to download the data.},
which is the first no-bias, high sensitive large-scale \COi/\COii/\COiii\ survey toward the Galactic plane,
somewhat solves such problem.
As one of the target regions of the survey,
a Galactic region of $l=[139.75,149.75]^{\circ}$, $b=[-5.25,5.25]^{\circ}$ (hereafter the G140 Region)
has been completely covered by nearly 2 yr of observation.
The total observing area is 105-square-degree,
and this paper takes a 96-square-degree part
(cf. Figure \ref{fig:map-bw} for the used part)
to study the spiral arm structure.
Another 9-square-degree part has been used for another study which will be published soon
(Xiong et al. 2017, to be accepted).

The G140 Region is located in the second quadrant of the MW,
which is a better place to study the arm structure
since the kinematic distance here is the monotonic function of LSR velocity.
Starting from $V_\mathrm{LSR} \sim 0$ km s$^{-1}$, increasingly negative velocities successively trace
the Local Arm, the Perseus Arm, the Outer Arm
and a new segment of spiral arm discovered by \citet{Sun2015ApJL} (hereafter New Arm).
In most MW spiral arm models,
the Perseus Arm, the Outer Arm, and the New Arm all comprise the major spiral arm of the MW
(cf. \citealt{SC2010}; \citealt{Sun2015ApJL}).
The Local Arm was once thought to be a spur structure,
until recently, when \citet{Xu2013ApJ,Xu2016} provided strong observational evidences to prove that
it is a larger structure, such as a branch.

Interestingly, since the G140 Region is just the intersection point of the Gould Belt and the Galactic plane
(cf. Figure 2 in \citealt{Grenier2004}),
the Local Arm in this place is made up of two layers
--- the Gould Belt layer which is associated with the Lindblad Ring traced by \HI\ gas
(\citealt{Lindblad1967BAN}; \citealt{Strauss1979A&A}),
and the Cam OB1 layer which is associated with the Cam OB1 association
(\citealt{Digel1996ApJ}; \citealt{Str2008}). 
The Cam OB1 layer is a star-forming active layer
that locates the famous young stellar object GL 490 and
two Sharpless H{\scriptsize~II} regions,
as well as the Cam OB1 association.
The star-forming activities in this region have been studied in detail by \citet{Str2007BaltA} and their series of works.
But the molecular cloud structure of this region was rarely studied.
Except for a small part studied by \citet{Digel1996ApJ},
up to now there has been no systematic CO observation completely covered this region.

In this paper, we report the \COi/\COii/\COiii\ mapping observation toward the G140 Region.
We mainly focus on studying the molecular structure of Local Arm and Perseus Arm here.
The study of star forming-activity will be presented in our next paper.
Section \ref{sec:obs} presents the CO observation condition and archival data of atomic hydrogen.
Section \ref{sec:para} presents the physical parameters of Local Arm and Perseus Arm derived by CO data.
Combing \HI\ data and part of Outer Arm results \citep{Du2016ApJS},
Section \ref{sec:arm} discusses the arm structures in the G140 Region.
Section \ref{sec:fila} presents five filamentary giant molecular clouds identified in the Perseus Arm
and discusses their relations to the MW large-scale structure.
Finally, Section \ref{sec:sum} presents the summary.

\section{Observation} \label{sec:obs}
\subsection{CO Observation}\label{subsec:COobs}
The \COi\ ($1-0$), \COii\ ($1-0$) and \COiii\ ($1-0$) lines 
were observed using the Purple Mountain Observatory 
Delingha (PMODLH) 13.7 m telescope from 2013 September to 2015 December
as one of the scientific demonstration regions for the MWISP project.
The three lines were observed simultaneously with the nine-beam superconducting array receiver (SSAR)
working in sideband separation mode
and with the fast Fourier transform spectrometer (FFTS) employed \citep{Shan2012IEEE}.
The \COi\ line is at the upper sideband (USB),
and the \COii\ and \COiii\ lines are at the lower sideband (LSB). 
Both of the bandwidths are 1000 MHz with 16,384 channels.
With on-the-fly (OTF) observing mode,
an area of $l=[139.75,149.75]^{\circ}$, $b=[-5.25,5.25]^{\circ}$ (105-square-degree in total) was covered,
and a 96-square-degree part is used in this work.
All the data were sampled every $30''$.
For the \COi\ observations (@ USB),
the main beam width was about $49''$, the main-beam efficiency ($\eta_\mathrm{MB,USB}$) was about 0.46,
and the typical rms noise level was about 0.5 K corresponding to a channel width of 0.16 km s$^{-1}$;
For the \COii\ and \COiii\ observations (@ LSB),
the main beam width was about $51''$, the main-beam efficiency ($\eta_\mathrm{MB,LSB}$) was about 0.48,
and the typical rms noise level was about 0.3 K corresponding to a channel width of 0.17 km s$^{-1}$.
All the data were corrected by $T_\mathrm{MB}=T^{*}_\mathrm{A}/\eta_\mathrm{MB}$.

\subsection{Archival Data of Atomic Hydrogen}\label{subsec:HIdata}
The 21 cm line data were retrieved from the Canadian Galactic Plane Survey (CGPS; \citealt{Taylor2003AJ}).
We downloaded data of $l=[140,150]^{\circ}$, $b=[-3,5]^{\circ}$
from the Canadian Astronomy Data Centre\footnote{http://cadc.hia.nrc.ca}.
The velocity coverage of the data is in the range of -153 to 40 km s$^{-1}$ with a channel separation of 0.82 km s$^{-1}$.
The survey has a spatial resolution of $58''$, which is comparable to our CO observations.

\section{Physical Parameters} \label{sec:para}

\subsection{Slicing the Region}\label{subsec:slice}
\begin{figure}[h]
\epsscale{1.2}
\plotone{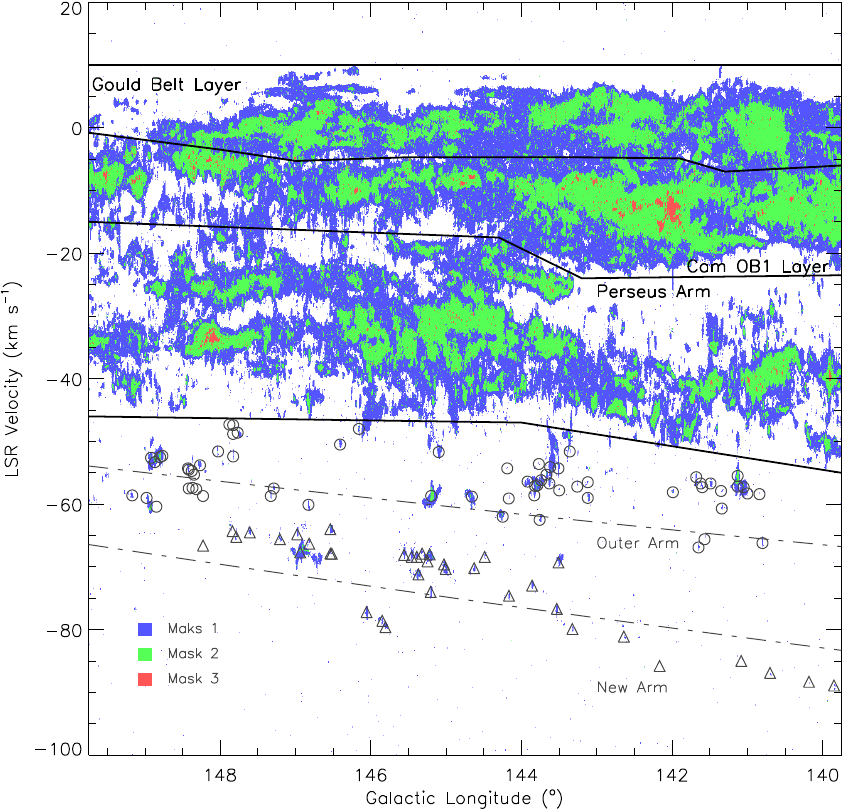}
\caption{\label{fig:lv-bw}
Longitude-velocity map of the whole area.
The blue, green, and red colors indicate Mask 1, Mask 2, and Mask 3, respectively.
(The meanings of the masks are presented in Section \ref{subsec:slice}.)
The Gould Belt layer, Cam OB1 layer, and Perseus Arm are divided by solid lines.
The circles and triangles present the Outer Arm clouds (identified by \citealt{Du2016ApJS})
and the New Arm clouds (identified by \citealt{Sun2015ApJL} and \citealt{Du2016ApJS}).
The two dashed lines indicate the projections of the Outer Arm spiral (fitted by \citealt{Du2016ApJS})
and the New Arm spiral (fitted by \citealt{Sun2015ApJL}), respectively.
}
\end{figure}

Figure \ref{fig:lv-bw} presents the LSR velocity distribution of the G140 Region.
The \COi, \COii\ and \COiii\ FITS cube data of the whole region
were integrated over all latitudes with 3$\sigma$ thresholds
(namely, \COi\ emission $\lesssim$ 1.5 K and \COii/\COiii\ emission $\lesssim$ 0.9 K are not integrated).
Then three high signal-to-noise ratio $l-V$ maps of those molecules are obtained.
Since the velocity resolutions are not the same (0.16 \kms @ \COi\ and 0.17 \kms @ \COii/\COiii), 
the velocity dimensions of the \COii\ and \COiii\ $l-V$ maps have been interpolated
for comparison with the \COi\ data.
Then, we define 
Mask 1 as the area where only \COi\ emission exists,
Mask 2 as the area where \COi\ and \COii\ emission both exist but \COiii\ emission does not,
and Mask 3 as the area where \COi, \COii\ and \COiii\ emission all exists.
The blue, green, and red colors in Figure \ref{fig:lv-bw} indicate Mask 1, Mask 2, and Mask 3, respectively.
Obviously, from LSR velocity $\sim$ 10 to $-50$ \kms\
the Gould Belt layer, Cam OB1 layer, and Perseus arm are successively located.
The black lines in Figure \ref{fig:lv-bw} outline their boundaries.
In addition, some Outer Arm and New Arm structure can also be seen in this map because of the high-quality data.
In order to show those two arms more clearly,
the molecular clouds and arm spiral projections
(identified and fitted by \citealt{Sun2015ApJL} and \citealt{Du2016ApJS}, 
see the caption of Figure \ref{fig:lv-bw} for more details) are plotted.

\begin{figure*}
\epsscale{1.2}
\plotone{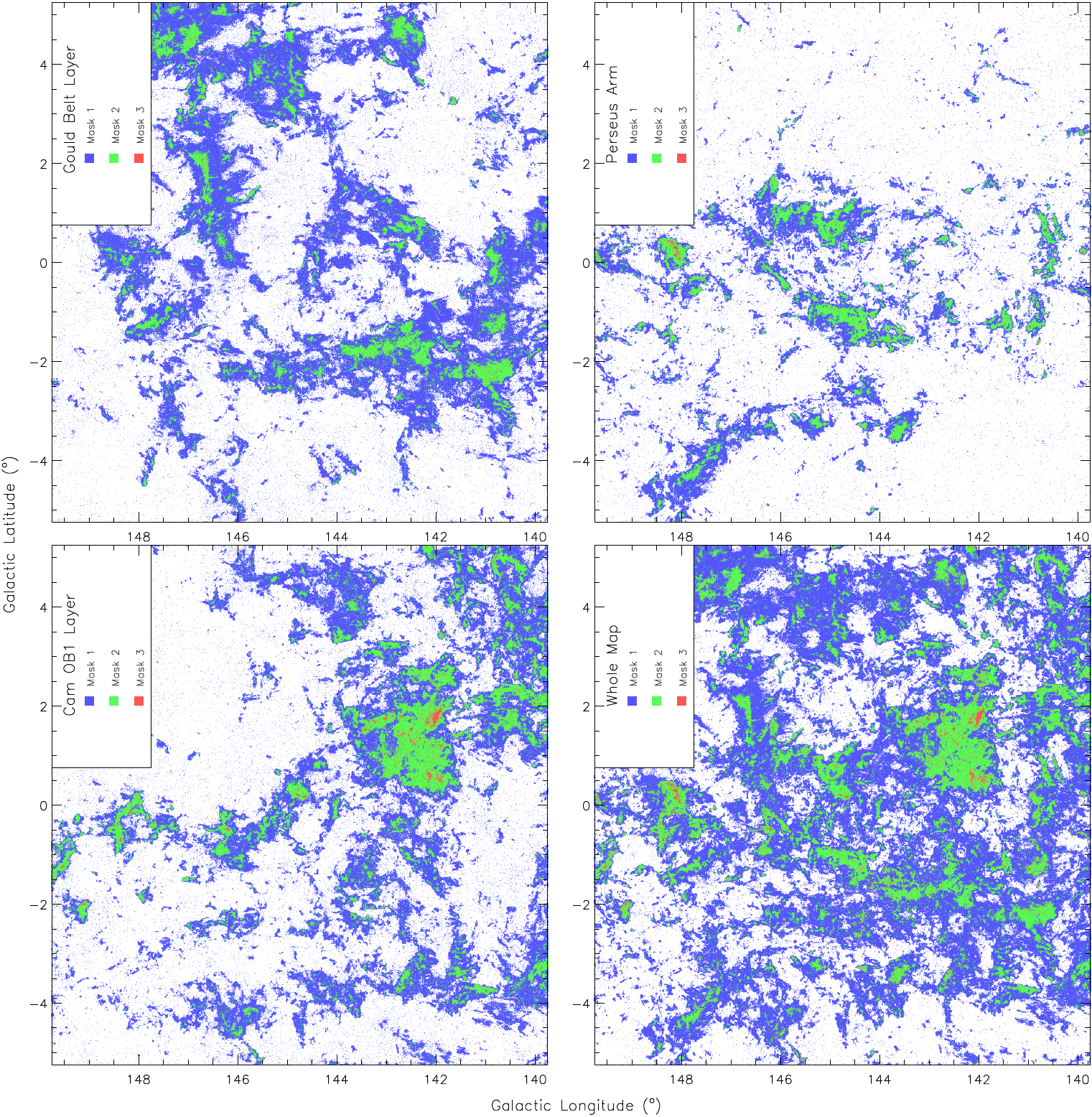}
\caption{\label{fig:map-bw}
Mask maps of the Gould Belt layer (top left panel),
the Cam OB1 layer (bottom left panel),
the Perseus Arm (top right panel)
and the whole area (bottom right panel).
The blue, green, and red color indicate Mask 1, Mask 2, and Mask 3, respectively.
(The meanings of the masks are presented in Section \ref{subsec:slice}.)
Note that the data in the upper left corner of each panel
(where the insert is located) were not used in this paper,
since they have been used for another study that will be published soon (Xiong et al. 2017, to be accepted).
}
\end{figure*}

According to the $l-V$ map, the FITS cube of the three molecules is integrated
over the velocity ranges, which are shown as the black lines in Figure \ref{fig:lv-bw}.
Also the integrated thresholds are 3$\sigma$.
Then the integrated intensity maps of the three molecules are obtained.
We also define Mask 1, Mask 2, and Mask 3 of the integrated maps,
and the definitions of the masks are the same as in the above presentation.
Figure \ref{fig:map-bw} shows the final results.
It can be seen that lots of molecular gas presents filamentary structure, especially in the Perseus Arm
(this will be discussed in Section \ref{sec:fila}).
In addition, \COiii\ is not rich in this region:
it is rarely distributed on the Gould Belt layer,
some is concentrated on the position of $(l,b)\sim(140,2)$\deg\ (where locates the GL 490) on the Cam OB1 layer, 
and on the Perseus Arm some is distributed on the position of $(l,b)\sim(148,0)$\deg.
This roughly indicates that the gas is relatively less dense in this region.

The distance of the Local Arm in this region has many photometric results.
The distance of the Gould Belt layer is in the range of 160 -- 300 pc (\citealt{Str2001A&A}; \citealt{Zd2005BaltA}),
and the Cam OB1 layer is about 1.0 kpc far away from us
(\citealt{Hum1978ApJS}; \citealt{Lyder2001AJ}; \citealt{Str2007BaltA}).
Although there is no photometric distance of the Perseus Arm in this region,
the nearby high-mass star-forming regions W 3OH (i.e. G133.94+01.06) and S Per (i.e. G134.62-02.19)
have the trigonometric parallax results,
which are 1.95 kpc \citep{Xu2006Sci} and 2.42 kpc \citep{Asaki2010ApJ}, respectively.
In addition, according to the Perseus Arm spiral fitted by \citet{Reid2014ApJ},
at the positions of $(l,b)=(140,0)$\deg\ and $(l,b)=(150,0)$\deg,
the distances are calculated to be 2.0 and 2.2 kpc, respectively.
Based on all the distance results presented above,
we finally adopt 200 pc, 1.0 kpc, and 2.1 kpc
as the distances of the Gould Belt layer, Cam OB1 layer, and Perseus arm, respectively.

\subsection{Physics of Layers}\label{subsec:phyLayer}

Assuming that \COi\ is optically thick, following \citet{Bourke1997ApJ},
the excitation temperature (\Tex) in each pixel of each layer can be calculated by
\begin{equation}
T_{\mathrm{ex}}=\frac{h\nu_{12}/k}{\ln(1+\frac{h\nu_{12}/k}{T_{\mathrm{MB,^{12}CO}}+\frac{h\nu_{12}/k}{\exp(h\nu_{12}/T_{\mathrm{bg}}-1)}})}
\end{equation}
where $h$ is the Planck constant, $k$ is the Boltzmann constant, $\nu_{12}$ is the frequency of \COi,
$T_{\mathrm{MB,^{12}CO}}$ is the peak main-beam temperature in each pixel,
and $T_{\mathrm{bg}}$ is the temperature of the cosmic microwave background.
The equation can be simplified as
\begin{equation}\label{eq:Tex}
T_{\mathrm{ex}}=\frac{5.532}{\ln(1+\frac{5.532}{T_{\mathrm{MB,^{12}CO}}+0.819})} \quad \mathrm{K}
\end{equation}
when all the constants are substituted.

\begin{figure}[h]
\epsscale{1.2}
\plotone{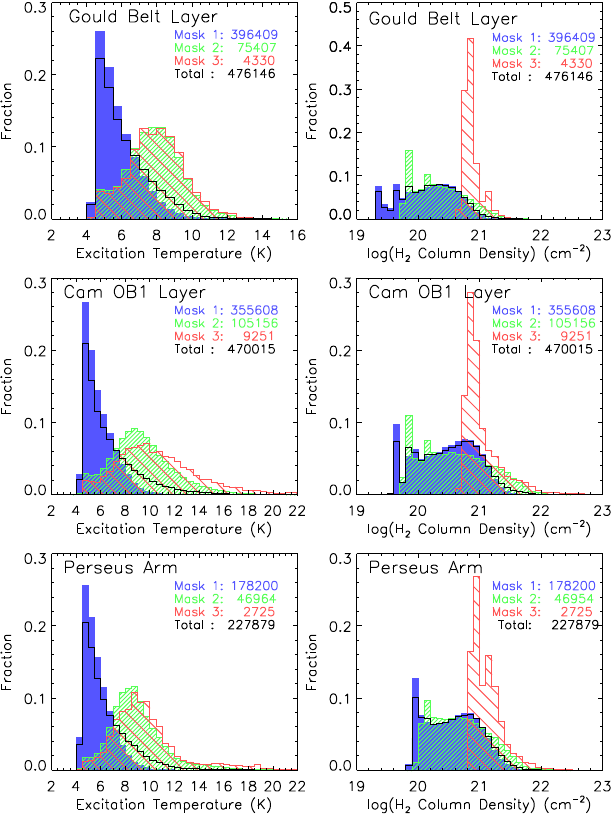}
\caption{\label{fig:N_T_dis_3mask}
Distributions of \Tex\ and \H2 column density of each layer.
The blue, green, red, and black colors in each panel indicate
the distributions of Mask 1, Mask 2, Mask 3, and the three masks in total, respectively.
The numbers with corresponding colors in each panel represent the pixel numbers.
}
\end{figure}

The three left panels of Figure \ref{fig:N_T_dis_3mask} show the distributions of \Tex\ in the three layers.
It can be seen that the excitation temperatures in this region are relatively low.
In the Cam OB1 layer, most of the \Tex\ of Mask 3 are in the range of 6 -- 16 K,
which are slightly higher than the other two layers (\Tex\ $\sim$ 6 -- 12 K).
The reason may be that the Cam OB1 layer has more star-forming activities.
Moreover, in each layer the \Tex\ peak distributions of Mask 1, Mask 2, and Mask 3 are in the increasing trend. 
This may indicate that the region where \COiii\ exists is relatively hotter
and the region where only \COi\ exists is relatively colder.

The \H2 column densities of the three masks are calculated by two different methods.
For Mask 1, where only \COi\ is detected, we use the X-factor to estimate the column density (hereafter the X-factor method):
\begin{equation}\label{eq:NH2-xfactor}
N_{\mathrm{Mask1,H_{2}}}=X\int T_{\mathrm{MB,^{12}CO}}dV
\end{equation}
where $\int T_{\mathrm{MB,^{12}CO}}dV$ is the integrated intensity of \COi\ in each pixel,
and the factor X is adopted from the results of \citet{Abdo2010ApJ},
namely, $0.87\times10^{20}$ cm$^{-2}$ (K $\cdot$ km s$^{-1}$)$^{-1}$ in the Gould Belt layer,
$1.59\times10^{20}$ cm$^{-2}$ (K $\cdot$ km s$^{-1}$)$^{-1}$ in the Cam OB1 layer,
and $1.9\times10^{20}$ cm$^{-2}$ (K $\cdot$ km s$^{-1}$)$^{-1}$ in the Perseus Arm.

For Mask 2 and Mask 3, the local thermodynamic equilibrium (LTE) condition is assumed.
Then, assuming that \COii\ and \COiii\ are optically thin and they share the same \Tex\ with \COi,
following \citet{Bourke1997ApJ} and \citet{Pineda2010ApJ},
the \COii\ column density of Mask 2 in each pixel can be calculated by
\begin{equation}\label{eq:N-13CO}
N_{\mathrm{Mask2,^{13}CO}}=2.42\times10^{14}\frac{\tau _{13}}{1-e^{-\tau_{13}}}
\frac{\int T_{\mathrm{MB,^{13}CO}}dV}{1-e^{-5.29/T_{\mathrm{ex}}}} \quad \mathrm{cm^{-2}}
\end{equation}
and for Mask 3 the \COiii\ column density in each pixel can be calculated by
\begin{equation}\label{eq:N-C18O}
N_{\mathrm{Mask3,C^{18}O}}=2.42\times10^{14}\frac{\tau _{18}}{1-e^{-\tau_{18}}}
\frac{\int T_{\mathrm{MB,C^{18}O}}dV}{1-e^{-5.27/T_{\mathrm{ex}}}} \quad \mathrm{cm^{-2}}
\end{equation}
where $\tau_{13}$ and $\tau_{18}$ respectively indicate the peak optical depth of \COii\ and \COiii,
$\int T_{\mathrm{MB,^{13}CO}}dV$ and $\int T_{\mathrm{MB,C^{18}O}}dV$
respectively indicate the integrated intensities of \COii\ and \COiii\ in units of K $\cdot$ \kms,
and \Tex\ is in units of K.
And $\tau_{13}$ and $\tau_{18}$ are calculated by
\begin{equation}
\tau_{13}=-\ln[ 1-\frac{T_{\mathrm{MB,^{13}CO}}}{5.29}([e^{5.29/T_{\mathrm{ex}}}-1]^{-1}-0.16)^{-1} ]
\end{equation}
and
\begin{equation}
\tau_{18}=-\ln[ 1-\frac{T_{\mathrm{MB,C^{18}O}}}{5.27}([e^{5.27/T_{\mathrm{ex}}}-1]^{-1}-0.17)^{-1} ]
\end{equation}
Where $T_{\mathrm{MB,^{13}CO}}$ and $T_{\mathrm{MB,C^{18}O}}$ are the peak main-beam temperatures
of \COii\ and \COiii, respectively.
Finally, according to the abundances of \COii\ and \COiii\ (\citealt{Ferking1982ApJ}; \citealt{CL1995AA}),
the \H2 column density of Mask 2 and Mask 3 in each pixel can be calculated by
\begin{equation}\label{eq:NH2-13CO}
N_{\mathrm{Mask2,H_{2}}}=7\times10^{5}\times N_{\mathrm{Mask2,^{13}CO}}
\end{equation}
and
\begin{equation}\label{eq:NH2-C18O}
N_{\mathrm{Mask3,H_{2}}}=7\times10^{6}\times N_{\mathrm{Mask3,C^{18}O}}
\end{equation}
Hereafter this column density estimation method is called the LTE method for short in this paper.

\begin{deluxetable}{ccrcrr}
\tablecaption{Layer Parameter\label{tab:Layer}}
\tablecolumns{6}
\tabcolsep = 1pt
\tablehead{
\colhead{Layer} &
\colhead{Mask} &
\colhead{Pixel} &
\colhead{Pixel \tablenotemark{a}} &
\colhead{Area} & 
\colhead{Mass} \\
\colhead{Name} & 
\colhead{Name} &
\colhead{Number} &
\colhead{Percentage} & 
\colhead{(pc\sq)} & 
\colhead{(\Msun)}
}
\startdata
\multirow{4}{*}{Gould Belt} & Mask 1 & 396,409 & 83.3\% & 335 & 1,607 \\
                            & Mask 2 &  75,407 & 15.8\% &  64 &   503 \\
                            & Mask 3 &   4,330 &  0.9\% &   4 &    63 \\
                            \cline{2-6}
                            & Total  & 476,146 &        & 400 & 2,200 \\
\hline
\multirow{4}{*}{Cam OB1}    & Mask 1 & 355,608 & 75.6\% & 7,522 & 77,395 \\
                            & Mask 2 & 105,156 & 22.4\% & 2,224 & 42,464 \\
                            & Mask 3 &   9,251 &  2.0\% &   196 &  7,394 \\
                            \cline{2-6}
                            & Total  & 470,015 &        & 9,900 & 127,300 \\
\hline
\multirow{4}{*}{Perseus Arm} & Mask 1 & 178,200 & 78.2\% & 16,624 & 198,130 \\
                             & Mask 2 &  46,954 & 20.6\% &  4,380 &  82,472 \\
                             & Mask 3 &   2,725 &  1.2\% &    254 &   8,787 \\
                             \cline{2-6}
                             & Total  & 227,879 &        & 21,300 & 289,400 \\
\enddata
\tablenotetext{a}{Pixel Percentage is the ratio of the pixel number in one\\
mask to the total pixel number in three masks of each \\layer.}
\end{deluxetable}

\begin{figure}[h]
\epsscale{1.2}
\plotone{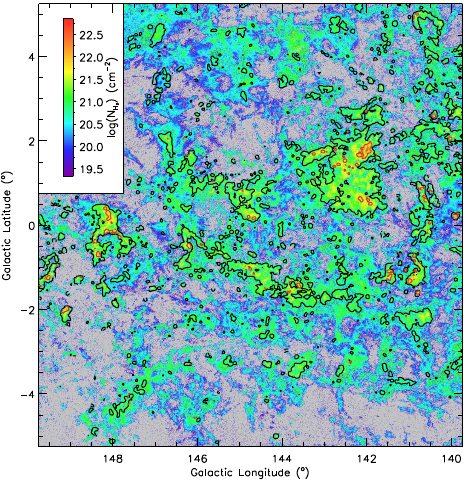}
\caption{\label{fig:Final_N}
\H2 column density of the whole area.
The black and red lines confine Mask 2 and Mask 3, respectively.
The rest of the area is Mask 1.
}
\end{figure}

Figure \ref{fig:Final_N} shows the \H2 column density result of the whole region.
Clearly, the \NH2 in the main part of the gas is about $10^{21}$ to $10^{21.5}$ cm$^{-2}$. 
The three right panels of Figure \ref{fig:N_T_dis_3mask} show the \H2 column density distribution of the three layers.
One can see that the \NH2 distribution of Mask 3 is much more concentrated (around a higher value of $10^{21}$ cm$^{-2}$)
than that of Mask 2 and Mask 3.
This indicates that the region traced by \COiii\ is relatively denser.

Knowing the \NH2 in each pixel, we can then estimate the mass of each mask:
\begin{equation}\label{eq:mass}
M_{\mathrm{Mask}}=2\mu m_{\mathrm{H}} a^{2}d^{2}\Sigma N_{\mathrm{Mask,H_{2}}}
\end{equation}
where $\mu$ (=1.36, \citealt{Hil1983QJRAS}) is the mean atomic weight per H atom in the ISM,
$m_{\mathrm{H}}$ is the H atom mass,
$a$ (=30$''$) is the angular size of each pixel,
$d$ is the distance of each layer (see Section \ref{subsec:slice}),
and $\Sigma N_{\mathrm{Mask,H_{2}}}$ refers to the \NH2 summation of all the pixels in each mask.
Table \ref{tab:Layer} shows the final results of pixel number,
pixel percentage
(the ratio of pixel number in one mask to the total pixel number in three masks),
physical area 
(the physical area is estimated by ``one pixel physical area'' $\times$ ``pixel number''),
and mass of each mask in each layer.
It is noticeable that the pixel percentage of Mask 2 and Mask 3 the in Cam OB1 layer is higher than that of other two layers.
This is perhaps because the Cam OB1 layer have more star-forming activities.

\section{Arm Structure} \label{sec:arm}
\subsection{Gas Distribution}\label{subsec:gasDis}

Most works about the gas distribution of the MW are mainly focused on the global and radial structure
(e.g., \citealt{Burton1975ApJ}; \citealt{Sod1987ApJ};
\citealt{NS2003PASJ} and their serial papers;
\citealt{Dua2015MNRAS}).
The distribution along the arm spiral direction is also meaningful.
For example, in the study of external galaxies by \citet{La2006ApJ},
the feather structure is well associated with the gas peak surface density
(see Figure 19 -- 20 of their paper).
Their work presented a good way of thinking about studying the substructure of our MW.
Substructures such as the branch, spur, and feather will be discussed in detail in Section \ref{sec:fila}.
In this section we just focus on the gas distribution along the Galactic longitude.

\begin{figure*}[t]
\epsscale{1.2}
\plotone{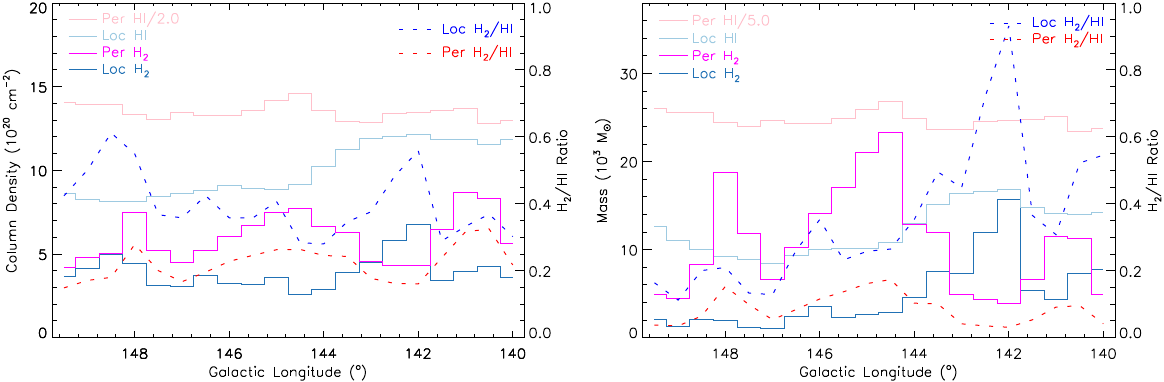}
\caption{\label{fig:N_M_alon}
Column density (left panel) and mass (right panel) distributions
of \H2 and \HI\ gases along the Galactic longitude.
Note that the column density and mass of \HI\ gas are divided by 2 and 5, respectively.
}
\end{figure*}

Figure \ref{fig:N_M_alon} shows the distributions of \HI\ and \H2 gases of
the Local Arm and Perseus Arm along the Galactic longitude.
Since the Local Arm shares two layers, the \NH2 in each pixel of this arm
is obtained by calculating the mean value in the same angular position of the Gould Belt layer and the Cam OB1 layer,
while the \H2 mass is obtained by adding together the data of the two layers.
Then, we calculate the mean value of \NH2 every 0.2 Galactic longitude degree of
the Local Arm and the Perseus Arm, respectively.
Thus the \NH2 distribution is obtained. 
And for the \H2 mass distribution, we calculate the summation of \H2 mass every 0.2 Galactic longitude degree.
To obtain the parameters of \HI\ gas,
we first slice the data into three layers just as the same LSR velocity ranges of Section \ref{subsec:slice}.
Second, assuming that \HI\ is optically thin, the surface density of each layer is calculated by
$\Sigma_{\mathrm{HI}}=1.82\times10^{18}m_{\mathrm{H}}\int T_{\mathrm{MB,HI}}d \,V$,
where $\int T_{\mathrm{MB,HI}}d \,V$ is the integrated intensity of \HI.
And then the \HI\ column density is obtained by the relation of
$\mathrm{1 \ M_{\odot}pc^{-2}=1.25\times10^{20} \ \ cm^{-2}}$.
Third, combining the distance, the mass of \HI\ in each pixel can be easily obtained,
and the calculation method is the same as the one in Section \ref{subsec:phyLayer}.
Finally, the \HI\ column density and mass distributions along the Galactic longitude are obtained
just using the same approaches of \H2 that have been presented above,
and the column density and mass ratios of \H2 and \HI\ gas are simply calculated by division.

\begin{deluxetable}{ccccc}
\tablecaption{Gas Mass of Spiral Arm \label{tab:ArmGas}}
\tablecolumns{5}
\columnsep=2pt
\tablehead{
\colhead{Spiral} &
\colhead{\H2} &
\colhead{\HI} &
\colhead{\H2 + \HI} &
\colhead{\H2 / \HI \tablenotemark{a}} \\
\colhead{Arm} & 
\colhead{(10$^{4}$\Msun)} &
\colhead{(10$^{4}$\Msun)} &
\colhead{(10$^{4}$\Msun)} & 
\colhead{}
}
\startdata
Local   & 9  &  25  & 34  & 0.36 \\
Perseus & 21 & 247  & 268 & 0.08 \\
Outer \tablenotemark{b}   & 15 & 624  & 639 & 0.02 \\
\enddata
\tablenotetext{a}{This column lists the mass ratio of \H2 to \HI.}
\tablenotetext{b}{The Outer Arm parameters are obtained from \citet{Du2016ApJS}.}
\end{deluxetable}

In Figure \ref{fig:N_M_alon} one can see that the gases in both the two arms vary smoothly, especially the \HI\ gas.
In addition, the \H2 column densities of the Local Arm and Perseus Arm are similar,
and the \H2 mass of the Perseus Arm is a little higher than that of the Local Arm.
However, the \HI\ gas exhibits a largely different performance.
Both the column density and mass of \HI\ gas on the Perseus Arm are much higher than those on the Local Arm.
Thus, the ratio of \H2 to \HI\ (hereafter the \H2/\HI) of the Local Arm is higher than that of the Perseus Arm.
This condition is also clearly shown in Table \ref{tab:ArmGas},
which summarizes the gas masses of the Local Arm and Perseus Arm
and also lists the masses of Outer arm in G140 Region for comparison
(the Outer Arm data are obtained from \citealt{Du2016ApJS}).
Since both the Outer Arm and Perseus Arm are the major spiral arm of MW
and the Outer Arm is located farther away from the Galactic center
(see Section \ref{sec:intro}),
it may be reasonable that \H2/\HI\ of the former is lower than that of the latter.
However, that the Local Arm --- as a subarm of the MW --- has
the highest \H2/\HI\ is somewhat abnormal.
Maybe it indicates a tendency that
the \H2 gas is more compactly located around the Galactic center than that of \HI\ gas.
However, another probable reason should NOT be ignored ---
that the \HI\ gas on the Local Arm is largely underestimated.
This will be presented in Section \ref{subsec:warpFlare}

\subsection{Warp and Flare}\label{subsec:warpFlare}

Warp is a common phenomenon among disk galaxies \citep{Binney1992ARA&A}.
The external galaxies with optical warps have long been recognized (\citealt{Sandage1961}; \citealt{Arp1966}),
but they were just thought to be some special cases since only a few were observed at that time.
Using \HI\ gas observation data, \citet{Sancisi1976A&A} found that
four out of five galaxies were gaseous warped.
Thereafter more gas warps were identified in a large number of external galaxies
(e.g., \citealt{Huch1980A&A}; \citealt{GR2002A&A}).
As it deserved, from that time on warp was regarded with special attention.
With no exception, our MW is also a warped galaxy.
In fact, the \HI\ gas warp was observed very early on
(\citealt{Burke1957AJ}; \citealt{Kerr1957AJ}; \citealt{Kerr1957Nature}; \citealt{Westerhout1957}).
The warp of our Galaxy has a large amplitude and is also asymmetric \citep{Levine2006ApJ}.
Combing recent \HI\ observation data,
\citet{Kalberla2009ARA&A} summarized a bended Galactic plane map (Figure 3 in their paper),
which showed a general distribution of Galactic warp.
Besides \HI\ gas, the MW warp has been observed and studied by many other tracers,
such as dust \citep{Fre1994ApJ}, CO \citep{Wouterloot1990A&A},
stars or OB stars (\citealt{Miyamoto1988A&A}; \citealt{Dehnen1998AJ}),
and IRAS point sources \citep{Djo989ApJ}.
Besides warp, flare --- as another gas behavior--- is also an interesting phenomenon among galaxies,
as well as our Galaxy.
In addition, the MW flare has been confirmed in both gaseous and stellar tracers
(e.g., \citealt{Momany2006A&A}; \citealt{Kalberla2009ARA&A}; \citealt{HD2015ARA&A}).
Usually flare and warp are discussed together
under the hypothesis that they are the response of dark matter
(e.g., \citealt{Binney1978MNRAS}; \citealt{Kalberla2007A&A}).
Unlike the difficult research situation in the MW spiral structure,
the study of warp and flare in the MW is relatively less troublesome
since the inclination problem that always exists among external galaxies can be discarded.
In this exact G140 Region,
we will morphologically present the warp and flare of both \HI\ and \H2 gases in the following paragraphs.

\begin{figure}[h]
\epsscale{1.0}
\plotone{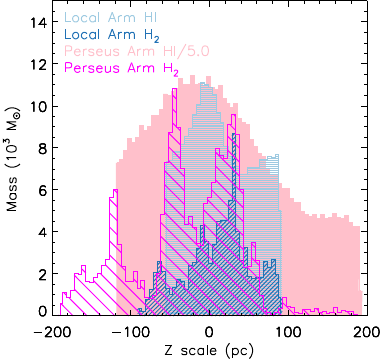}
\caption{\label{fig:Mass_dis}
Gas distribution of the Local Arm and Perseus Arm along the Z scale.
Note that the mass of \HI\ gas of the Perseus Arm is divided by 5.
}
\end{figure}

Figure \ref{fig:Mass_dis} shows the mass distributions of
\H2 and \HI\ gas on the Local Arm and Perseus Arm along the Z scale.
First, according to the Galactic latitude and distance, we calculate the Z scale of every pixel of each layer.
Second, we calculate the total mass of every 5 pc Z scale of each layer;
thus, we can get the gas distributions of 3 layers.
Then, we add the distributions of the Gould Belt layer and Cam OB1 layer as the Local Arm gas distribution,
and the distributions of the two spiral arms are finally obtained.
Obviously, the gas distribution is incomplete, due to the limited observation scope along the Galactic latitude. 
Some \H2 gas on the Local Arm seems to be not completely observed
--- the \H2 gas mass suddenly becomes zero at a Z scale of about 90 pc (where $b\sim5$\deg).
In addition, part of the \HI\ gas on the Perseus Arm and large amount of \HI\ gas on Local Arm are not completely observed.
The distributions suddenly stop at Z scales of about -120, -60, 90, and 200 pc, respectively. 
In other words, the total mass of \HI\ gas on the two spiral arms, especially on the Local Arm, is largely underestimated
(which is one of the probable reason that the \H2/HI of the Local Arm is largely higher than that of the Perseus Arm
mentioned in Section \ref{subsec:gasDis}).

\begin{deluxetable}{ccccc}
\tablecaption{Thickness and Height of Arm \label{tab:ArmHeight}}
\tablecolumns{5}
\tabcolsep = 10pt
\tablehead{
\colhead{Spiral} &
\multicolumn{2}{c}{Thickness} &
\multicolumn{2}{c}{Height} \\
\colhead{Arm} &
\colhead{\H2} &
\colhead{\HI} &
\colhead{\H2} & 
\colhead{\HI} \\
\colhead{} & 
\colhead{(pc)} &
\colhead{(pc)} &
\colhead{(pc)} &
\colhead{(pc)}
}
\startdata
Local   & 117  &  220  & 18  & -2 \\
Perseus & 149  &  291  & -16 & -19 \\
Outer \tablenotemark{a}  &  60  &  550  & 170 & 160 \\
\enddata
\tablenotetext{a}{The Outer Arm parameters are obtained from \citet{Du2016ApJS}.}
\end{deluxetable}

\begin{figure}[h]
\epsscale{1.0}
\plotone{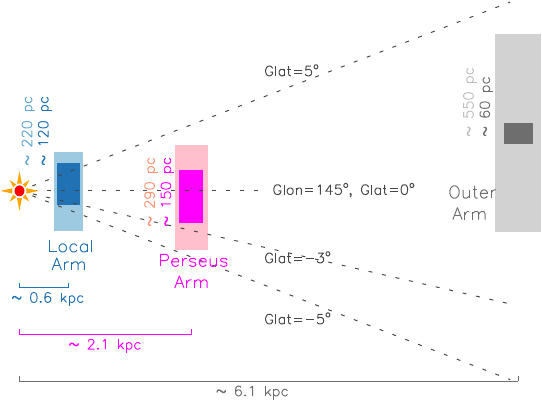}
\caption{\label{fig:cartoon}
Sketch map of the thicknesses and heights of three spiral arms.
The Galactic latitudes plotted are proportionally zoomed in.
The light-blue, pink, and light-gray rectangles indicate the \HI\ gas of the Local Arm, Perseus Arm,
and Outer Arm, respectively.
The dark-blue, magenta, and dark gray-rectangles indicate the \H2 gas of the three arms.
The numbers with corresponding colors beside the rectangles mark the thickness values of \HI\ and \H2 gases of the arms.
}
\end{figure}

Despite the incomplete Z scale gas distribution,
we can still estimate the thickness and height of each arm by fitting the distribution in a Gaussian curve.
We define the FWHM as the arm thickness and the centric position at Z scale as the arm height. 
Table \ref{tab:ArmHeight} lists the final results,
as well as the parameters of the Outer Arm in the G140 Region for comparison (data from \citealt{Du2016ApJS}).
A more direct perspective on the arm distributions can be seen in Figure \ref{fig:cartoon}.
We obtain the mean distance of the Gould Belt layer and Cam OB1 layer as the distance of the Local Arm.
For the Outer Arm distance,
we adopt the mean distance of all the Outer Arm clouds in the G140 Region with the cloud mass as weight. 
One can see that the distance between the Outer Arm and Perseus Arm
is much larger than that between the Perseus Arm and Local Arm,
which cannot be obviously shown on the $l-V$ map.
(However, the Outer Arm distance may be overestimated a little since the cloud distances adopted are kinematic distances;
details are presented in Section 4.2 of \citealt{Du2016ApJS}.)
In addition, both the Local Arm and the Perseus Arm are almost located at the Galactic plane,
while the Outer Arm lies as high as $\sim150-200$ pc above it.
This suggests that the MW warp in this region may start at a place between the Perseus Arm and Outer Arm.
In addition, the \HI\ gas flare is obvious, but it does not exist in \H2 gas.
The \H2 gas thickness of Perseus Arm is a little thicker than that of the Local Arm,
but it becomes much thinner on the Outer Arm.
Maybe one of the reasons is that on the edge of the galaxy some arms become narrow (e.g., \citealt{Honig2015ApJ}).
But another possible reason should not be ignored ---
that the Outer Arm molecular clouds have not been completely detected yet.

\section{The Filamentary Giant Molecular Clouds in the Perseus Arm} \label{sec:fila}

Filament, as an important star-forming stage \citep{Andre2014PP},
has obtained a lot of attention in recent years.
In the Gould Belt filament study of \citet{Andre2014PP},
the filament length is at a scale of $\sim1-10$ pc, while its scale width is $\sim0.1$ pc.
But in the larger-scale, there also exist some filamentary cloud structures.
The first large scale filamentary infrared dark cloud (IRDC), ``Nessie'', was identified by \citet{Jackson2010ApJ},
which is $\sim80$ pc long and $\sim0.5$ pc wide.
Then, \citet{Goodman2014ApJ} added that ``Nessie'' might probably be as long as 430 pc.
In addition, they related it to the MW structure and called it the ``bone'' of the Galaxy.
Subsequently, more filamentary clouds with such a large scale were identified,
and their relations with the large-scale structure of MW were also under research
(e.g., \citealt{Li2013AA}; \citealt{Tack2013AA}; \citealt{BB2014ASSP}; \citealt{Su2016ApJ}).

As mentioned in Section \ref{subsec:slice},
one can clearly see that in Figure \ref{fig:map-bw} the molecular gas presents filamentary structure,
especially in the Perseus Arm.
The molecular cloud complex of the Perseus Arm in this region was only partly studied by \citet{Digel1996ApJ} before.
In this section,
we have totally identified 5 filamentary giant molecular clouds (FGMCs) in the Perseus Arm,
among which 4 are newly identified.
Their properties, comparisons with ``Nessie'' and other large-scale filamentary clouds,
and their relations to the MW structure will be presented in the following.

\subsection{General Properties}\label{subsec:filaGeneral}

\begin{figure}[t]
\epsscale{1.2}
\plotone{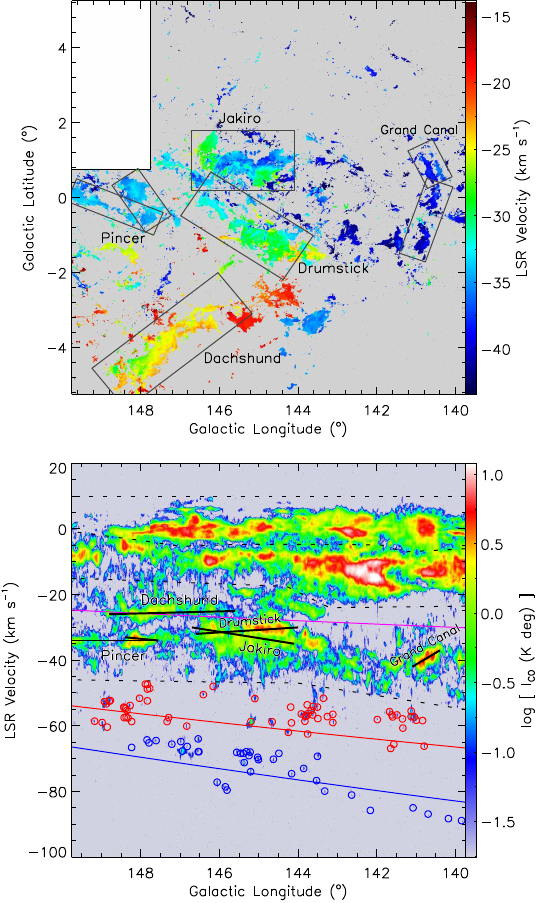}
\caption{\label{fig:filadis}
Distribution of the FGMCs on the Perseus Arm.
The top panel shows the locations of FGMCs on the \COi\ velocity-coded map of the Perseus Arm.
The bottom panel shows the FGMCs on the longitude-velocity map.
The thick black lines present the FGMCs.
The black dashed lines divide the regions of the Gould Belt layer, Cam OB1 layer, and Perseus Arm.
The red and blue circles indicate the Outer arm clouds (identified by \citealt{Du2016ApJS})
and the New Arm clouds (identified by \citealt{Sun2015ApJL} and \citealt{Du2016ApJS}).
The red and blue lines presented the projections of the Outer Arm spiral (fitted by \citealt{Du2016ApJS})
and the New Arm spiral (fitted by \citealt{Sun2015ApJL}), respectively.
The magenta line indicates the projection of the Perseus Arm spiral fitted by \citet{Reid2014ApJ}.
}
\end{figure}

Based on the spatial morphology and velocity continuity of \COi\ and \COii\ data,
we totally identified 5 FGMCs in the Perseus Arm,
and named them ``Grand Canal'',``Jakiro'',``Drumstick'',``Pincer'', and ``Dachshund'', respectively.
Among them, ``Grand Canal'' has been partly detected by \citet{Blitz1982ApJS} 
and studied by \citet{Digel1996ApJ} before
(more details are presented in Section \ref{subsub:GrandCanal}),
while the other 4 clouds are newly identified.
In addition, ``Pincer'' is a cloud that consists of two filamentary structures
(``Pincer-longer'' and ``Pincer-shorter'').
Figure \ref{fig:filadis} presents the distribution of the FGMCs
(see the caption for more details).
Table \ref{tab:FGMC} lists the properties of FGMCs,
including the centric position,
the height to the Galactic plane,
the angle between the longer side of FGMC and the Galactic plane (hereafter FP angle),
the mean and maximum excitation temperature,
the dense gas mass function (DGMF),
and the parameters traced or derived by \COi, \COii\, and \COiii, respectively.
Among them, the \COi\ and \COii\ rows list the LSR velocity range,
the length and width of the FGMC, the ratio of length to width, the physical area, the mean and maximum \H2 column density,
and the mass derived by \COi\ and \COii\, respectively,
while the \COiii\ rows only list the similar parameters derived by \COiii\
except length, width, and the length-to-width ratio,
since the \COiii\ distribution in the FGMC is patchy rather than filamentary.
Note that (i) the ``Pincer-longer'' also lacks the parameters of
length, width, and length-to-width ratio traced by \COii\ for the similar reason,
and (ii) the parameters traced by \COiii\ of ``Dachshund'' are not listed since nearly no \COiii\ emission is detected.

\begin{figure*}[t]
\epsscale{1.2}
\plotone{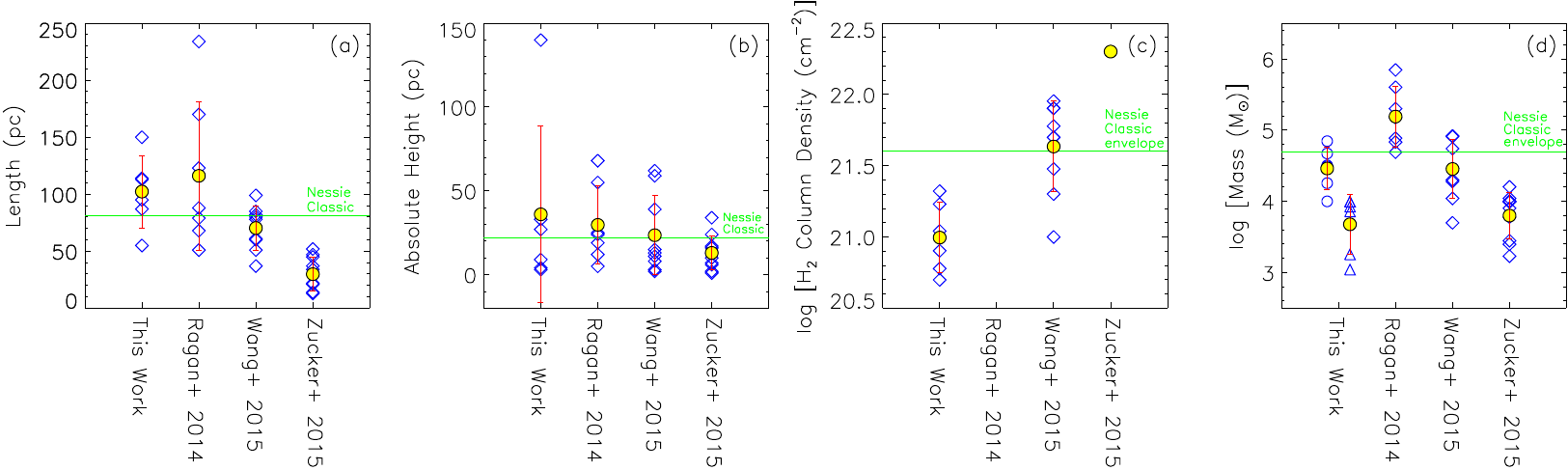}
\caption{\label{fig:compareOtherFila}
Comparisons of the (a) length, (b) absolute height,
(c) mean \H2 column density, and (d) mass of the FGMCs with other works.
The yellow filled circle indicates the mean value of each work.
The standard deviation is marked by the red error bar.
The green line in each panel indicates the value of ``Nessie'' (\citealt{Jackson2010ApJ}; \citealt{Goodman2014ApJ}),
in panel (a) and (b) it indicates the value of ``Nessie Classic'',
and in panel (c) and (d) it indicates the value of the envelope of ``Nessie Classic'' traced by HNC.
Note that:
(1) In panel (c), the \H2 column densities of ``This Work'', ``Wang+ 2015'', and ``Zucker+ 2015''
are from \COii\ derived result, SED fitting result, and an assumed value, respectively;
(2) In panel (d), The blue circles and triangles of ``This work'' indicate the masses
derived by \COi\ and \COii, respectively.
In addition, the masses of ``Ragan+ 2014'', ``Wang+ 2015'', and ``Zucker+ 2015''
are derived by \COii\ data, SED fitting column density, and assumed column density, respectively.
}
\end{figure*}

The estimation of all the physical parameters is similar to the method presented in Section \ref{sec:para}.
The only difference is that we do not define masks in this section.
Namely,
the area where we calculate \COi-derived mass also includes the area where \COii\ and \COiii\ emission exists,
and similarly for the area where we calculate \COii-derived mass.
The calculation methods for all the parameters are as follows.

First, we respectively integrate the \COi, \COii, and \COiii\ FITS cube of each FGMC
over the velocity range that is listed in Table \ref{tab:FGMC}.
The integrated threshold is also $3\sigma$,
and the pixels without signals in the final integrated map are set to be NULL.
Then, we adopt ``the total number of pixels that are not NULL'' $\times$ ``the angular area of each pixel''
as the total angular area.
According to the distance (2.1 kpc), the physical area can be easily estimated.
Second, the length is adopted as the physical length of the longer side of the box that confines the FGMCs
shown in the top panel of Figure \ref{fig:filadis}.
The width is obtained by dividing the physical area by the physical length.
Then, the length-to-width ratio can be calculated accordingly.
Third, the excitation temperature is calculated by the \COi\ peak main-beam temperature,
and the calculation method is the same as in Equation \ref{eq:Tex}.
Fourth, the \H2 column density is also estimated by two methods.
For the area traced by \COi, the X-factor method is adopted,
and the calculation method is the same as in Equation \ref{eq:NH2-xfactor};
for the area traced by \COii\ and \COiii, the LTE method is adopted,
and the calculation method is the same as in Equation \ref{eq:N-13CO} -- \ref{eq:NH2-C18O}.
Fifth, the calculation method of mass traced by different CO molecules is the same as in Equation \ref{eq:mass}.
Finally, we consider \COiii\ tracing the dense gas
and define the DGMF as the ratio of the molecular mass derived by \COiii\ to the one derived by \COii.

Generally,
on morphology, these FGMCs are 
(i) at the scale length of $\sim100$ pc,
and (ii) at the scale width of $\sim10$ pc and $\sim1$ pc traced by \COi\ and \COii\, respectively.
On distribution,
(i) except for ``Dachshund'', the other FGMCs all distribute near the Galactic plane (the absolute hight $\sim10$ pc),
and (ii) the FP angle range is as large as from $\sim0$\deg to $\sim90$\deg;
On physical property,
(i) their excitation temperature is similar ($\sim10$ K),
(ii) the DGMF range is large ($\sim0\%-9\%$),
(iii) 3 FGMCs have relatively obvious velocity gradients (details in Section \ref{subsec:filaInd}),
(iv) the mean \H2 column densities are not high
(the densities derived by \COi, \COii\, and \COiii\ are all $\sim10^{21}$ cm$^{-2}$),
and (v) the masses derived by \COi\ are all at the giant molecular cloud (GMC) scale.

The properties of these FGMCs are somewhat different from those of the giant filamentary IRDC ``Nessie''.
Their lengths are similar to ``Nessie Classic'' (namely, the part identified by \citealt{Jackson2010ApJ}),
but their widths are $\sim10-100$ times larger than that of ``Nessie''.
In addition, their mean \H2 column densities are lower than that of ``Nessie'' for one order of magnitude.
However, the properties between the FGMCs traced by \COii\
and the envelope of ``Nessie Classic'' traced by HNC \citep{Goodman2014ApJ} are not very different:
their width scale is similar,
and also they share the same \H2 column density scale of $10^{21}$ cm$^{-2}$.

In recent years, using the infrared and molecular line data, 
\citet{Ragan2014AA}, \citet{Wang2015MNRAS} and \citet{Zuker2015ApJ}
have all systematically identified and studied several giant filamentary clouds.
Figure \ref{fig:compareOtherFila} shows the comparisons 
of length, absolute height (namely, the absolute distance to the Galactic plane), \H2 column density, and mass
between this work and theirs as well as ``Nessie''.
Detailed description about the figure is presented in the note.
One can see that the lengths, masses derived by \COi\ and the absolute heights of this work are similar to those of others,
while the \H2 column densities are relatively lower than in other works.

\subsection{Individuals}\label{subsec:filaInd}

\subsubsection{``Grand Canal''} \label{subsub:GrandCanal}

The FGMC ``Grand Canal'' is named after a famous canal in China --- the Beijing--Hangzhou Grand Canal,
which is the longest and one of the oldest man-made rivers in the world.
Unlike other large rivers in China, the Beijing--Hangzhou Grand Canal is a south--north direction river,
similar to the FGMC ``Crand Canal'', which is almost vertical to the Galactic plane (FP angle $\sim81$\deg).
Using a CO tracer, a point position of this FGMC was first detected by \citet{Blitz1982ApJS} with the name of BFS27.
In the study of \citet{Digel1996ApJ}, ``Crand Canal'' was completely detected and was divided into 3 parts
(numbers of 50 -- 52 in their catalog,
and BFS27 is located at Number 52).
Here using relatively highly sensitive and high-resolution data, we consider those three clouds as one filamentary structure
and study it again in this section.

Figure \ref{fig:Per-F1_map} presents the properties of this FGMC.
From panel (a), one can see that the \COiii\ distribution is clumpy.
These \COiii\ clumps are located at $(l,b)\cong(140.6,0.6)$\deg, $(140.5,-0.2)$\deg, $(140.9,-0.7)$\deg,
and $(141.0,-1.3)$\deg, where the excitation temperature and \H2 column density are relatively higher,
as shown in panel (c), (d) and (e).
Among them, at the position of $(l,b)\cong(141.0,-1.3)$\deg
(which is also the position near BFS27), the \COiii\ main-beam temperature is the highest,
and the three molecular line profiles at this position are shown in panel (f).
In addition, the FGMC breaks at the position of $(l,b)\sim(140.75,-0.4)$\deg,
and bends at the position of $(l,b)\sim(140.5,0.4)$\deg.
North of $b\sim0.4$\deg, the FGMC splits into two parallel parts with different velocity components,
which can be seen in panels (b) and (g).
Moreover, from panels (b) and (g), one can see that the FGMC presents a weak velocity gradient along the major axis.
In panel (g), at the position of 0\deg\ its velocity $\sim-41$ \kms,
while at the position of 3.3\deg\ the velocity becomes to $\sim-37$ \kms.
Thus, we can roughly estimate that the velocity gradient is $\sim0.04$ \kms\ pc$^{-1}$.

Compared to other FGMCs,
(i) its \COi\ length-to-width ratio is the highest;
(ii) its \H2 column densities derived by \COi, \COii, and \COiii\ are all the highest;
(iii) its excitation temperatures are the highest;
and (iv) it is the only FGMC of which the \COiii\ distribution is clumpy.

\subsubsection{``Jakiro''} 

\begin{figure}[h]
\epsscale{1.2}
\plotone{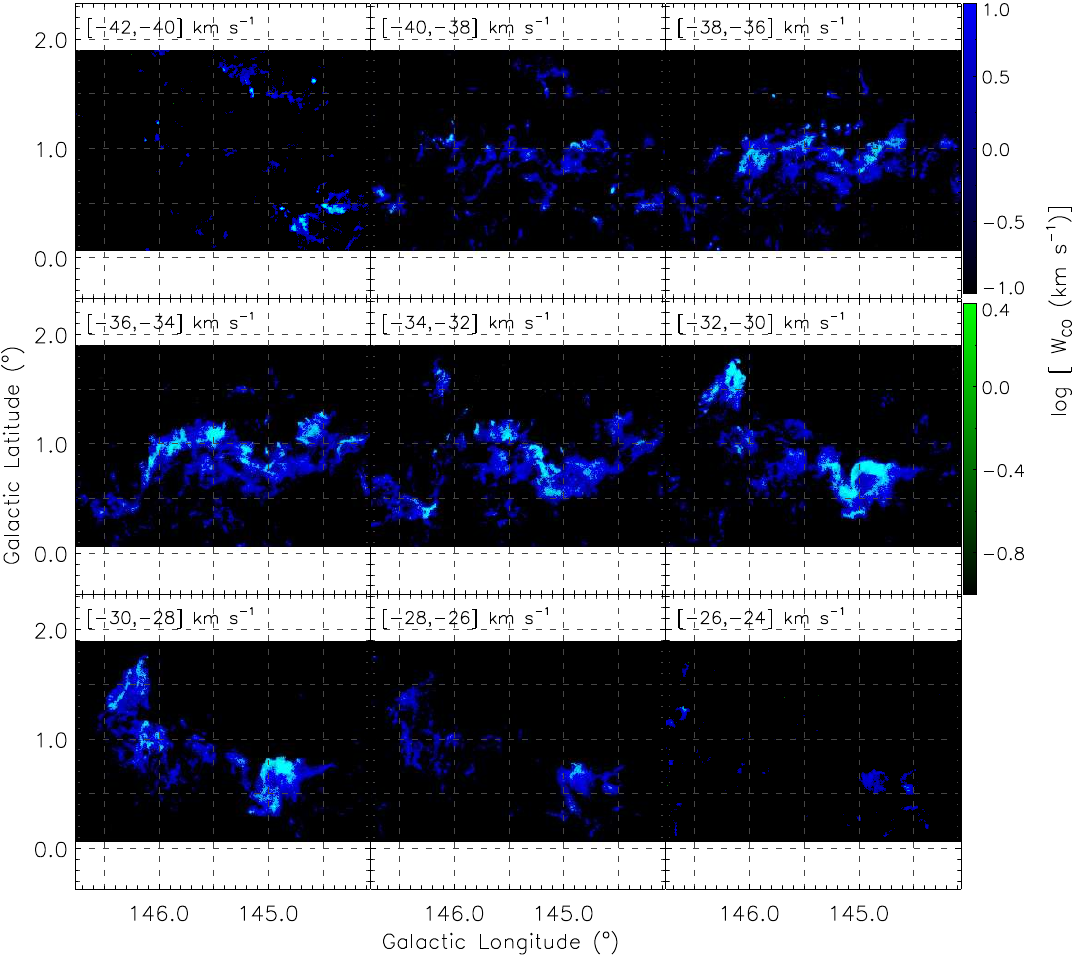}
\caption{\label{fig:Per-F2_channelmap}
Channel map of \COi\ (blue) and \COii\ (green) of FGMC ``Jakiro''.
Note that the integrated intensity value is logarithmic.
}
\end{figure}

\begin{figure*}
\epsscale{1.0}
\plotone{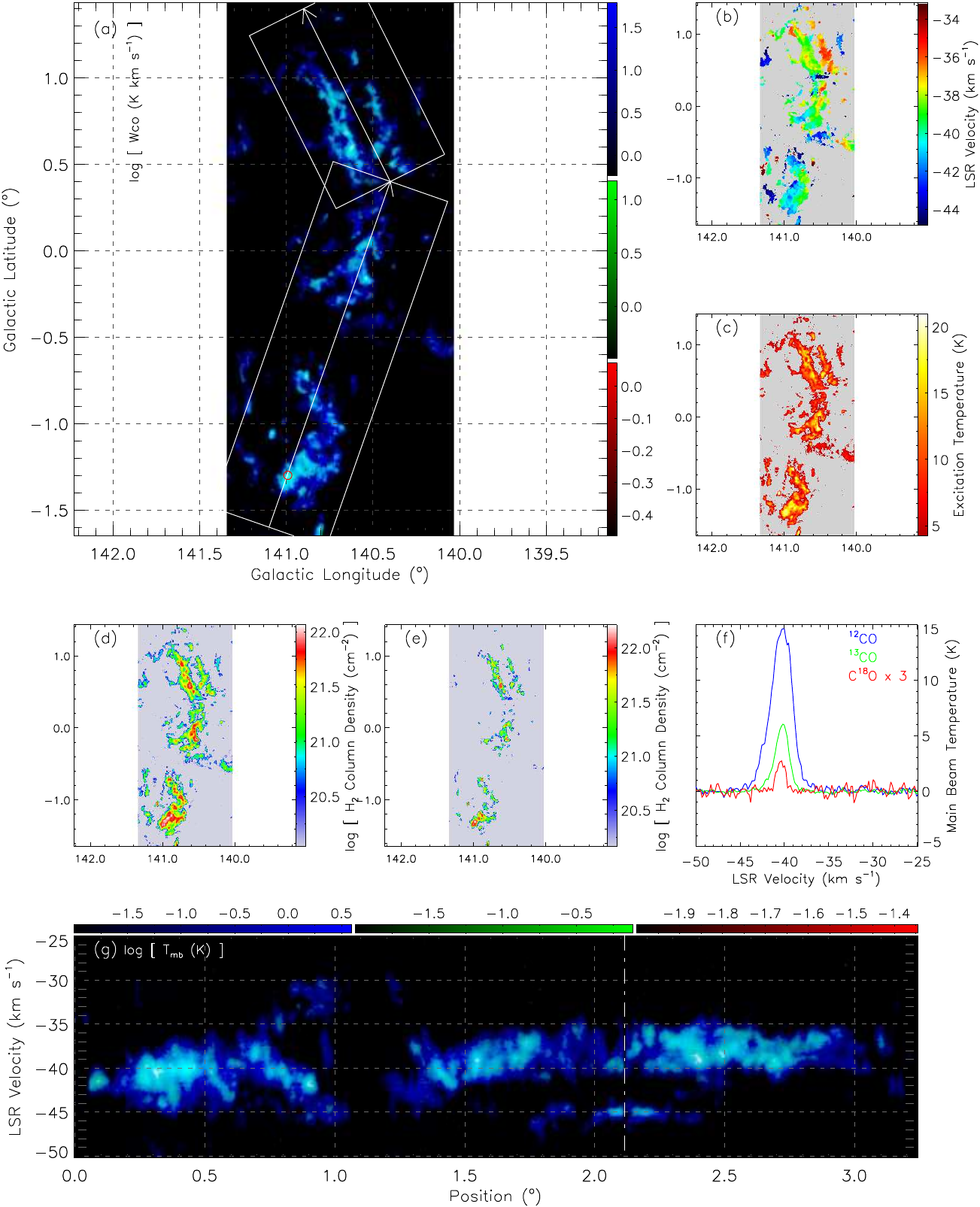}
\caption{\label{fig:Per-F1_map}
Properties of the FGMC ``Grand Canal''.
(a) is the integrated intensity map of \COi\ (blue), \COii\ (green) and \COiii\ (red).
Note that the value is logarithmic.
The red circle indicates the location where \COiii\ peak main-beam temperature is the highest.
(b) is the \COi\ velocity-coded map.
(c) is the excitation temperature map derived by \COi\.
(d) and (e) are \H2 column density map derived by \COi\ and \COii, respectively.
Note that the value is logarithmic.
(f) is the averaged spectrum of \COi\ (blue), \COii\ (green) and \COiii\ (red) of 9 pixels of the red circle in panel (a).
Note that the \COiii\ spectrum is multiplied by 3.
(g) is the average velocity-position map of the white rectangles in panel (a),
and the white arrows in panel (a) indicate the position direction.
Note that the white pecked line in panel (g) refers to the turning point in panel (a).
}
\end{figure*}

\begin{figure*}
\epsscale{1.0}
\plotone{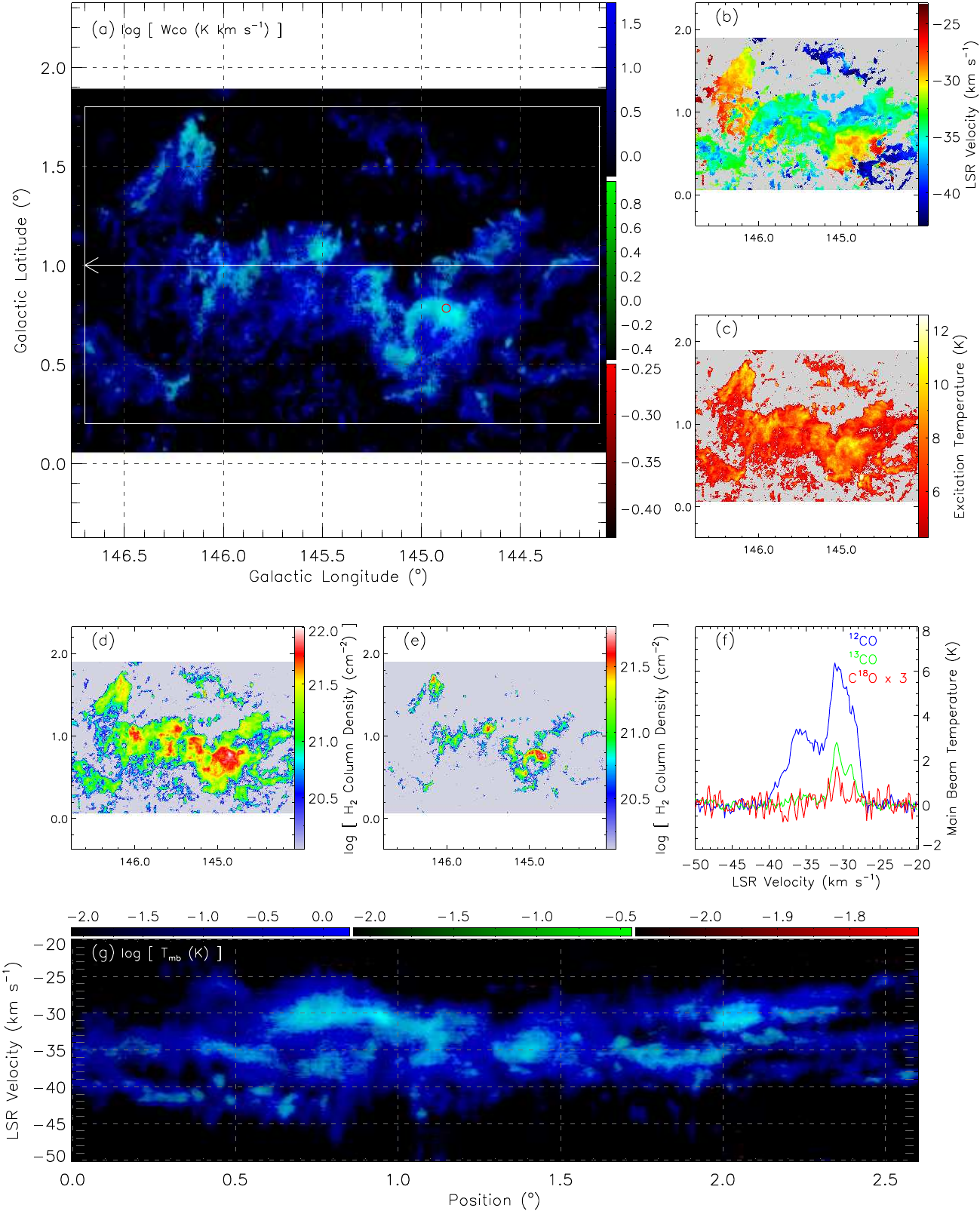}
\caption{\label{fig:Per-F2_map}
Properties of the FGMC ``Jakiro''. The meaning of each panel is consistent with Figure \ref{fig:Per-F1_map}.
}
\end{figure*}

The FGMC ``Jackiro'' is named after its complicated velocity components,
just as shown in panels (b), (f), and (g)  of Figure \ref{fig:Per-F2_map}.
Figure \ref{fig:Per-F2_channelmap} shows its channel map.
One can see that this FGMC seems to be composed of two ``S'' shapes that twist together,
while the \COii\ seems to be mainly distributed in one of the ``S'' shapes,
just as shown in panel (e) of Figure \ref{fig:Per-F2_map}.
In addition, \COiii\ distribution is very diffused.

Compared to other FGMCs,
(i) it is the only one which is located above the Galactic plane;
(ii) its \COi\ length to width ratio is the lowest;
and (iii) its physical areas and masses traced and derived by \COi\ and \COii\ are both the largest.

\subsubsection{``Drumstick''} 

\begin{figure*}
\epsscale{1.0}
\plotone{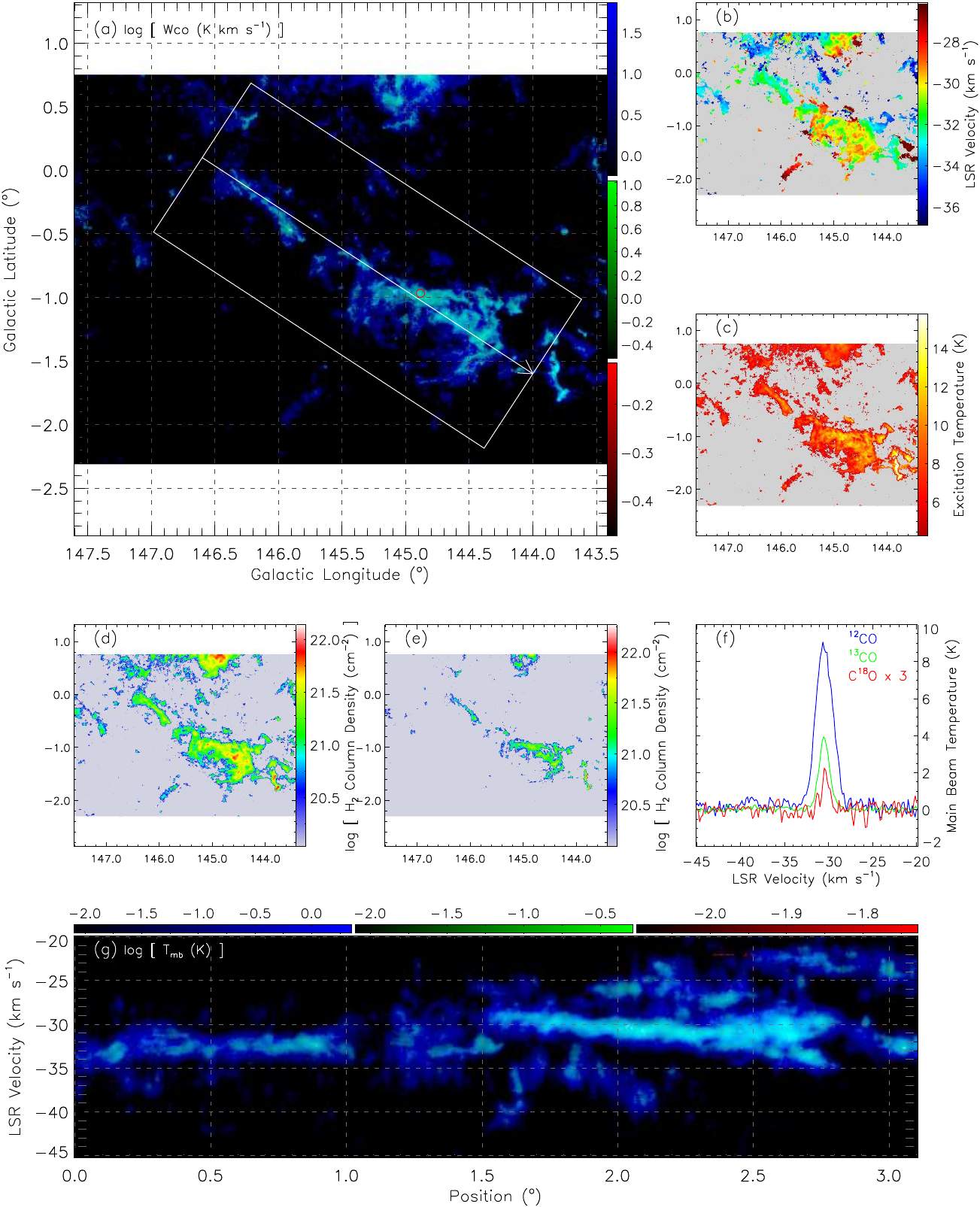}
\caption{\label{fig:Per-F3_map}
Properties of the FGMC ``Drumstick''. The meaning of each panel is consistent with Figure \ref{fig:Per-F1_map}.
}
\end{figure*}

The FGMC ``Drumstick'' is named after its shape
--- one narrower part at the northeast side and one wider part at the southwest side,
as shown in Figure \ref{fig:Per-F3_map}.
These two parts break at the position of $(l,b)\sim(145.8,-0.4)$.
The wider part contains more complicated velocity components.
As shown in panel (g) of Figure \ref{fig:Per-F3_map},
the wider part is composed of about four velocity components:
the main component is at the major axis with velocity $\sim-30$ \kms,
and the other three components distribute at both sides of the main one
with velocity $\sim-35$, $-28$, and $-23$ \kms, respectively.
In addition, this FGMC presents a weak velocity gradient along the major axis.
At the position of 0\deg, its velocity $\sim-33$ \kms,
and at the position of 3.1\deg, the velocity becomes $\sim-30$ \kms.
Thus, the velocity gradient is $\sim0.03$ \kms\ pc$^{-1}$.
The excitation temperature and \H2 column density of the wider part are more higher than
those of the narrower part. 
The \COiii\ distribution is relatively diffused in this FGMC,
and almost all the \COiii\ distributes at the wider part.

\subsubsection{``Pincer''} 

\begin{figure*}
\epsscale{1.0}
\plotone{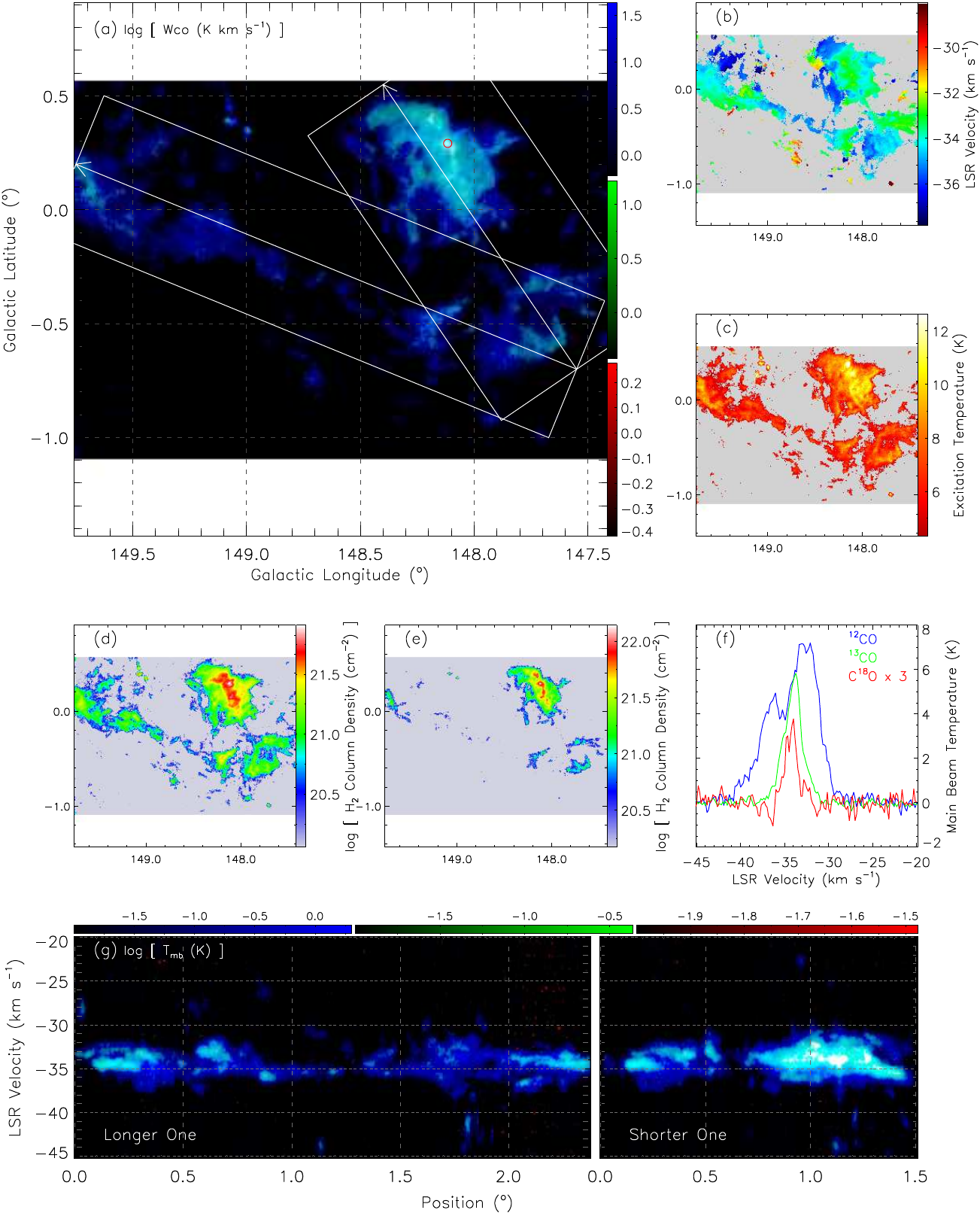}
\caption{\label{fig:Per-F4_map}
Properties of the FGMC ``Pincer''. The meaning of each panel is consistent with Figure \ref{fig:Per-F1_map}.
Note that in panel (g), the left map and right map refer to the velocity-position maps of
``Pincer-longer'' and ``Pincer-shorter'', respectively.
}
\end{figure*}

The FGMC ``Pincer'' consists of two filamentary structures ---
the ``Pincer-longer'' and the ``Pincer-shorter''.
Those two parts present very different properties, as shown in Figure \ref{fig:Per-F4_map}.
``Pincer-longer'' is very diffused.
The \COii\ on it is even too diffused to form a complete filamentary structure.
The \COiii\ is also rare.
The \H2 column density and excitation temperature are both very low.
On the other hand,
``Pincer-shorter'' is more compact,
and its \H2 column density and excitation temperature are both higher.
The \COiii\ is rich and largely concentrated in the vicinity of $(l,b)\sim(148.1,0.3)$\deg,
where its excitation temperature is relatively higher.

Compared to other FGMCs,
(i) this FGMC is the only one that consists of two filamentary components;
(ii) the \COii\ and \COiii\ on the ``Pincer-shorter'' are the richest;
and (iii) the DGMF of ``Pincer-shorter'' is the highest.

\subsubsection{``Dachshund''} 

\begin{figure*}
\epsscale{1.0}
\plotone{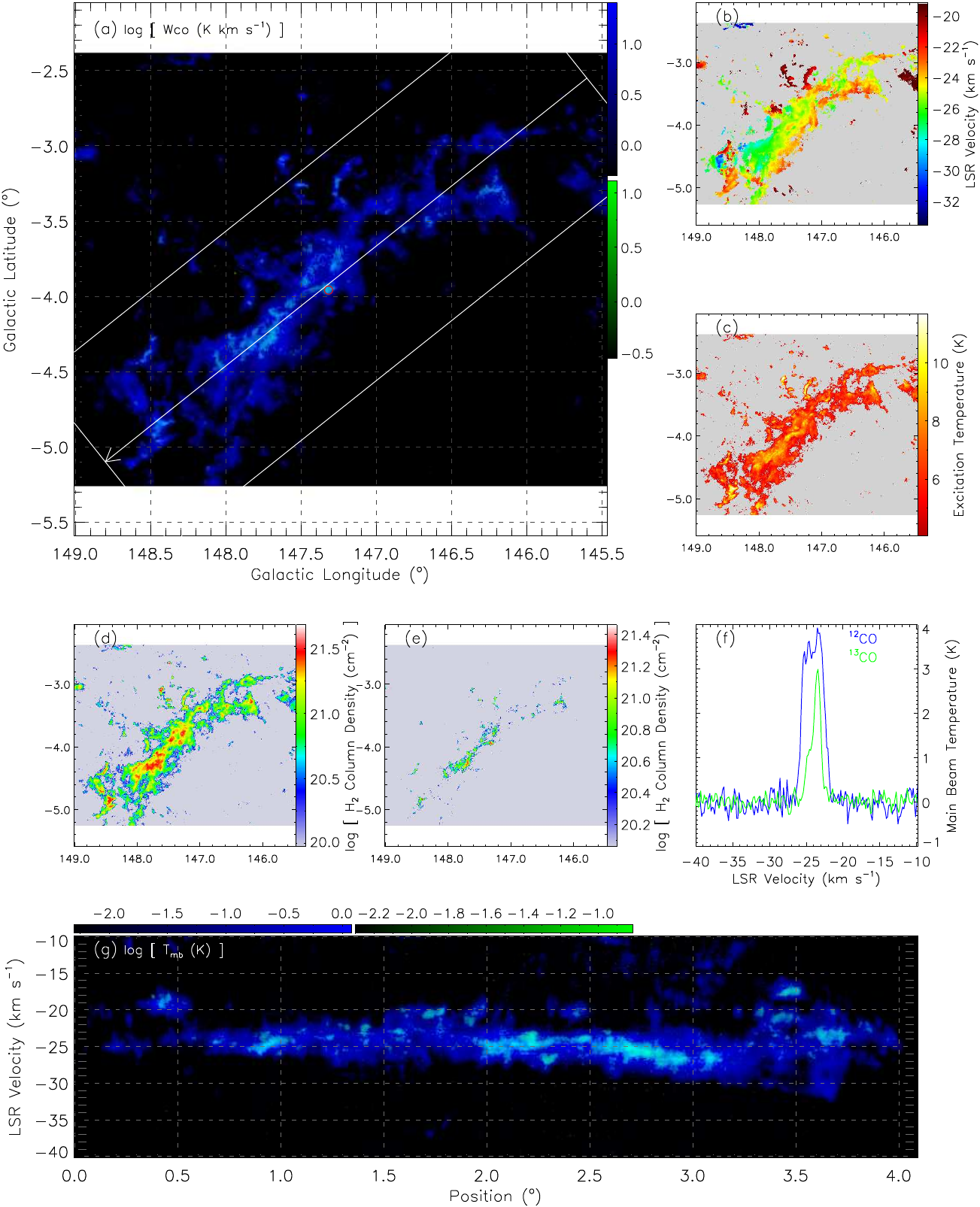}
\caption{\label{fig:Per-F5_map}
Properties of the FGMC ``Dachshund''. The meaning of each panel is consistent with Figure \ref{fig:Per-F1_map}.
Note that \COiii\ is not plotted since it is too weak to be detected.
And correspondingly the red circle in panel (a) indicates
the location where \COii\ peak main-beam temperature is the highest.
}
\end{figure*}

The shape of FGMC ``Dachshund'' is interesting.
It looks like a dog running from the northwest to the southeast, as shown in Figure \ref{fig:Per-F5_map}.
This FGMC is a very diffused cloud.
The \COiii\ is too weak to be detected.
In addition, its excitation temperature is also very low.
However, it is noticeable that along its minor axis it presents an obvious velocity gradient.
At the ``belly'' of the dog, the velocity is $\sim-23$ \kms,
and at the ``back'' of the dog, it becomes $\sim-27$ \kms.
Thus, the velocity gradient is estimated to be $\sim0.25$ \kms pc$^{-1}$.

Compared to other FGMCs,
(i) its height is the largest, namely, it is located farthest away from the Galactic plane;
(ii) it is the longest FGMC;
and (iii) its DGMF is $\sim0$ because no \COiii\ is detected.

\subsection{Relations between FGMCs and Galactic Structure}\label{subsec:largeStru}

The spiral arms of many galaxies are not smooth \citep{weaver1970}.
On the arms there exist the so-called substructures, such as the branch, spur, and feather
(e.g., \citealt{Lynds1970}; \citealt{Pidd1973ApJ}; \citealt{Elme1980ApJ}; \citealt{La2006ApJ}).
In the study of 7 galaxies, \citet{Elme1980ApJ} has found that
the width and length of the spur structure are at the order of magnitude of 100 pc and 1 kpc, respectively.
In fact, the substructures of our galaxy were also observed and studied early on
(e.g., \citealt{Sofue1976A&A}; \citealt{RC1979ApJ}).
Recently, a spur with kpc-length between the Local Arm and the Sagittarius Arm has been identified by \citet{Xu2016},
which further implies the complicated situation of the MW substructures.

Apparently, those substructures are also filamentary.
However, they are hugely longer and wider than the FGMCs and other giant filamentary clouds studied by
\citet{Ragan2014AA}, \citet{Wang2015MNRAS} and \citet{Zuker2015ApJ}.
In addition, up to now the longest giant filamentary cloud in the MW is just $\sim500$ pc long \citep{Li2013AA}.
Such FGMCs are indeed too small compared to the substructures like the spur.
On the other hand,
they are about 10 times larger than the traditional Gould Belt filaments studied by \citet{Andre2014PP}.
Table \ref{tab:filaComp} lists a comparison among several filamentary structures with different scales.
There seems to exist a delicate trend in angle (FP angle or pitch angle) such
that the angle range and value become small as the filamentary structure becomes large.
In addition, their locations also change from subarm to major arm and then to interarm.
As mentioned at the beginning of Section \ref{sec:fila},
the relations between such a giant filament and MW large scale-structure are under research.
Some subtle trends have indeed been found (e.g., \citealt{Ragan2014AA}),
but strong evidences still need more observational samples.

\begin{deluxetable}{ccccc}
\tablecaption{Comparison of Filamentary Structure\label{tab:filaComp}}
\tablecolumns{5}
\tabcolsep = 1pt
\tablehead{
\colhead{Structure} &
\colhead{Length} &
\colhead{Angle} &
\colhead{Location} &
\colhead{Ref}
}
\startdata
Gould Belt filament   & $1-10$ pc     & -            & sub-arm   & 1 \\
FGMC                  & $\sim100$ pc  & $0-90$\deg   & major arm & -  \\
Other giant filament  & $\sim100$ pc  & $0-70$\deg   & major\&inter arm & 2 \\
Feather               & $\sim1$ kpc   & $\sim50$\deg & major arm & 3,4 \\
Spur                  & $\sim1$ kpc     & $\sim60$\deg & inter-arm & 4,5 \\
Local arm             & $>5$ kpc             & $\sim11$\deg &    -    & 6,7 \\
Major arm of MW       & $>10$ kpc     & $7-20$\deg   & -       & 7,8 \\
\enddata
\tablecomments{
1: \citet{Andre2014PP}; \\
2: \citet{Ragan2014AA}, \citet{Wang2015MNRAS}, \\ and \citet{Zuker2015ApJ}; \\
3: \citet{Lynds1970}; \\
4: \citet{La2006ApJ}; \\
5: \citet{Elme1980ApJ}; \\
6: \citet{Xu2013ApJ,Xu2016}; \\
7: \citet{Reid2014ApJ}; \\
8: \citet{SC2010}, \citet{Sun2015ApJL}, \\ and \citet{Du2016ApJS}. \\
Note that the angle of ``FGMC'' and ``Other giant filament'' \\
refers to the angle between filamentary structure and the \\
Galactic plane; While the angle in other rows refers to the \\ pitch angle.
}
\end{deluxetable}

\section{Summary} \label{sec:sum}

The G140 Region, namely, the Galactic region of
$l=[139.75,149.75]^{\circ}$, $b=[-5.25,5.25]^{\circ}$,
is one of the target survey areas of the ongoing Milky Way Imaging Scroll Painting (MWISP) project.
The \COi\ ($1-0$), \COii\ ($1-0$) and \COiii\ ($1-0$) lines were observed simultaneously
using the Purple Mountain Observatory Delingha (PMODLH) 13.7 m telescope.
With nearly 2 yr of observations, this region has been completely covered.
We take a 96-square-degree part of this region to mainly study the molecular structure of the Local Arm and Perseus Arm.
Combining the \HI\ data and part of the Outer Arm results, the gas distribution, warp and flare are discussed.
In addition, five filamentary giant molecular clouds (FGMCs) on the Perseus Arm are identified,
and their relations to the MW large scale structure are discussed.
The main results of the G140 Region are as the follows:

(1) The Local Arm consists of two layers
--- the Gould Belt layer and the Cam OB1 layer.
Their molecular masses are
$\sim2.2\times10^{3}$ \Msun\ and $\sim12.7\times10^{4}$ \Msun,
respectively.
In total the Local Arm molecular mass is
$\sim12.9\times10^{4}$ \Msun.
The Perseus Arm molecular mass is
$\sim28.9\times10^{4}$ \Msun.

(2) The mass ratios of \H2 to \HI\ gas on the Local Arm, Perseus Arm, and Outer Arm are 0.36, 0.08, and 0.02, respectively.
However, the ratio of Local arm may be overestimated since the \HI\ gas on Local arm is largely underestimated.

(3) The \H2 gas thicknesses of Local arm, Perseus arm, and Outer arm are 117 pc, 149 pc, and 60 pc, respectively;
the \HI\ gas thicknesses of the three arms are 220 pc, 291 pc, and 550 pc, respectively.
The \H2 gas height of the three arms are 18 pc, -16 pc, and 170 pc, respectively;
the \HI\ gas height of the three arms are -2 pc, -19 pc, and 160 pc, respectively.
It can be seen that the warp structure of both atomic and molecular gas is obvious,
while the flare structure only exists in atomic gas.

(4) Five FGMCs with lengths $\sim 100$ pc on the Perseus Arm are identified,
among which four are newly identified.
Their masses derived by \COi, \COii, and \COiii\ are
$\sim10^{4}$ \Msun, $\sim10^{3}$ \Msun, $\sim10^{2}$ \Msun, respectively;
and their mean \H2 column densities derived by \COi, \COii\ and \COiii\ are all $\sim10^{21}$ cm$^{-2}$.

\acknowledgments
We are grateful to all the members of the Milk Way Scroll Painting survey group, 
especially the staff of Qinghai Radio Observing Station at Delingha for technical support.   
We also acknowledge the anonymous referee for the helpful comments.
This work is supported by the National Natural Science Foundation of China
(Grant Nos. and 11673066, 11133008, and 11233007)
and the Key Laboratory for Radio Astronomy.

\vspace{5mm}



\clearpage
\begin{deluxetable*}{l|lcc|C{1.65cm}|C{1.65cm}|C{1.65cm}|C{1.65cm}|C{1.65cm}|C{1.65cm}}
\tablecaption{Properties of FGMCs \label{tab:FGMC}}
\tablecolumns{10}
\tabcolsep = 1pt
\scriptsize
\tablehead{
\multicolumn{2}{l}{\multirow{2}{*}{Name}} &
\colhead{} &
\multicolumn{1}{c|}{\multirow{2}{*}{(1)}} &
\multicolumn{1}{c|}{Grand} &
\multicolumn{1}{c|}{\multirow{2}{*}{Jakiro}} &
\multicolumn{1}{c|}{\multirow{2}{*}{Drumstick}} &
\multicolumn{1}{c|}{Pincer} &
\multicolumn{1}{c|}{Pincer} &
\multicolumn{1}{c}{\multirow{2}{*}{Dachshund}} \\
\multicolumn{2}{l}{} &
\colhead{} &
\multicolumn{1}{c|}{} &
\multicolumn{1}{c|}{Canal} &
\multicolumn{1}{c|}{} &
\multicolumn{1}{c|}{} &
\multicolumn{1}{c|}{-longer} &
\multicolumn{1}{c|}{-shorter} &
\multicolumn{1}{c}{}
}
\startdata
\multicolumn{2}{l}{Glon} & (\deg) & (2) & 140.6 & 145.4 & 145.3 & 148.7 & 148.0 & 147.2  \\
\multicolumn{2}{l}{Glat} & (\deg) & (3) & -0.1  & 0.9   & -0.8  & -0.3  & -0.1  & -3.8   \\
\multicolumn{2}{l}{FP Angle} & (\deg) & (4) &   81 &     0 &    33 &    22 &    56 & 39     \\
\multicolumn{2}{l}{Height} & (pc) & (5) & -4    & 33    & -27   & -9    & -3    & -140   \\
\multicolumn{2}{l}{\Tex} & (K)    & (6) & 9/26  & 7/16  & 8/16  & 6/12  & 7/21  &  6/12  \\
\multicolumn{2}{l}{DGMF} & (\%)   & (7) & 3.2   & 0.1   & 0.8   & 0.7   & 8.7   & 0      \\
\hline
\multirow{6}{*}{\COi} & \Vlsr & (\kms) & (8) & [-45,-33] & [-43,-23] & [-37,-26] & [-38,-29] & [-38,-29] & [-34,-19] \\
                      & Scale & (pc$\times$pc) & (9) &
                      $113\times9$  & $95\times31 $ & $114\times19$  & $87\times10$  & $55\times15$ & $150\times16$ \\
                      & \multicolumn{2}{l}{Len/Wid}   & (10) & 13 & 3 & 6 & 9 & 4 & 9 \\
                      & Area & (pc\sq) & (11) & 991 & 2918 & 2115 & 848 & 847 & 2389 \\
                      & \NH2 & ($10^{21}$cm$^{-2}$) & (12) & 1.9/11.6 & 1.5/10.1 & 1.4/7.3 & 0.7/3.3 & 1.3/8.1 & 0.8/4.0 \\
                      & Mass & (10$^{4}$\Msun) & (13) & 3.1 & 7.0 & 4.7 & 1.0 & 1.8 & 3.2 \\
\hline
\multirow{6}{*}{\COii} & \Vlsr & (\kms) & (14) & [-45,-34] & [-41,-25] & [-36,-27] & [-37,-30] & [-37,-30] & [-30,-20] \\
                       & Scale & (pc$\times$pc) & (14) &
                       $113\times3$  & $95\times8$ & $114\times6$  & -  & $55\times7$ & $120\times2$  \\
                       & \multicolumn{2}{l}{Len/Wid}   & (15)  & 37 & 13 & 20 & - & 8 & 58 \\
                       & Area & (pc\sq) & (16) & 341 & 722 & 663 & 142 & 359 & 248  \\
                       & \NH2 & ($10^{21}$cm$^{-2}$) & (17) & 2.1/16.4 & 0.8/6.3 & 1.1/5.0 & 0.6/2.7 & 1.7/21.9 & 0.5/2.9 \\
                       & Mass & (10$^{3}$\Msun) & (18) & 11.4 & 9.8 & 11.5 & 1.4 & 10.0 & 2.1 \\
\hline
\multirow{4}{*}{\COiii} & \Vlsr & (\kms) & (19) & [-43,-36] & [-32,-28] & [-34,-29] & [-37,-31] & [-37,-31] & - \\
                        & Area & (pc\sq) & (20) & 4.3 & 0.3 & 2.1 & 0.2 & 12.9 & - \\
                        & \NH2 & ($10^{21}$cm$^{-2}$) & (21) & 5.3/12.8 & 2.3/2.6 & 2.8/4.7 & 2.7/3.1 & 4.2/12.2 & - \\
                        & Mass & (10$^{2}$\Msun) & (22) & 3.6 & 0.1 & 0.9 & 0.1 & 8.7 & -
\enddata
\tablecomments{
Row (1): the name of FGMC.
Row (2) -- (3): the central position of FGMC in Galactic coordinate.
Row (4): the angle between the longer side of FGMC and the Galactic plane.
Row (5): the height to the Galactic plane, which is calculated by height = distance $\times$ $\sin$(latitude).
Row (6): mean/max excitation temperature of FGMC.
Row (7): dense gas mass function of FGMC.
Row (8) -- (13): parameters traced or derived by \COi, respectively are
LSR velocity range, length $\times$ width, ratio of length to width,
area, mean/max \H2 column density, and mass.
Row (14) -- (18) and Row (19) -- (22): parameters traced or derived by \COii\ and \COiii, respectively.
And the meaning of each row is consistent with Row \COi.
Note that Row (14) -- (15) of ``Pincer-longer'' and Row (19) -- (22) of ``Dachshund'' are null because of the weak emission.
}
\end{deluxetable*}



\end{document}